# Defects of graphene on Ir(111): rotational domains and ridges

Elena Loginova, Shu Nie, Konrad Thürmer, Norman C. Bartelt, and Kevin F. McCarty∗

Sandia National Laboratories
Livermore, CA

**Abstract**

We use low-energy electron microscopy (LEEM), low-energy electron diffraction (LEED) and scanning tunneling microscopy (STM) to study different orientations of single-layer graphene sheets on Ir(111). The most-abundant orientation has previously been characterized in the literature. Using selective-area LEED we find three other variants, which are rotated 14°, 18.5° and 30° with respect to the most common variant. The ~30°-rotated structure is also studied by STM. We propose that all 4 variants are moiré structures that can be classified using simple geometric rules involving periodic and quasi-periodic structural motifs. In addition, LEEM reveals that linear defects form in the graphene sheets during cooling from the synthesis temperature. STM shows that these defects are ridges, suggesting that the graphene sheets delaminate locally as the Ir substrate contracts.

## 1. Introduction

For long time, carbon on surfaces has been a focus of many studies. Research on graphene, single-layer sheets of graphitic carbon, has increased recently because of the unusual and potentially useful electronic properties of this two-dimensional material.[1,2] One method of graphene synthesis employs single-crystal metal substrates as templates to grow graphene with few defects.[3] Typical templates include Ni(111),[4-7] Ru(0001),[8-17] Ir(111)[18-24] and Pt(111).[25-31] The latter three surfaces are not well lattice-matched to graphene. Because of this mismatch and the strong C-C bonding within the graphene sheet, the metal lattice and the graphene lattice form moiré structures.[3] Traversing the large unit cells of the moiré patterns, the position of the C atoms relative to the metal atoms changes. That is, the 6-membered C rings of graphene change from being centered over metal atoms to being centered over the hollow sites of the metal. These spatial changes in C/metal bonding cause the graphene sheets to be periodically buckled.[3]

The Cologne group has studied in detail the atomic structure of one moiré structure of graphene on Ir(111).[18, 19, 23, 24, 32] In this paper, we further examine the graphene/Ir(111) system and find three additional orientations of graphene. While less abundant than the structure previously characterized, the three additional structures can still occur as relatively large domains, tens of microns in spatial extent. We characterize these orientations of graphene using low-energy electron microscopy (LEEM) and selected-area low-energy electron diffraction (LEED). One of the new orientations is also characterized by scanning tunneling microscopy (STM). We find that the four types of graphene merely differ in their azimuthal orientation with respect to the Ir surface, similar to the previously reported rotational variants on Pt(111).[25, 28, 31] We propose atomic models for the three new rotational variants and contrast the structures with the structure of the previously reported variant. In addition, we show that graphene films on Ir(111) develop ridges during cooling from the synthesis temperature.

**2. Experimental methods**

Single-layer graphene was grown on Ir(111) in a LEEM. Details of the Ir surface preparation are reported elsewhere.[22] Graphene was grown from 3 C sources: impurity C segregating from the bulk of the crystal, C deposited from a graphite rod heated by an electron-beam, and decomposing ethylene gas of 99.999% purity. The Ir temperature was measured using a type-C thermocouple spot-welded to the side of the crystal. Graphene growth was directly observed by collecting time sequences of LEEM images. Once grown, the different rotational variants of graphene were further characterized in the LEEM using imaging to analyze their spatial distribution, measuring how the intensity of the specularly reflected electron beam changed with electron energy, and by collecting electron diffraction patterns from selected areas. For the latter, diffraction from a specific area on the surface is obtained using an aperture to limit the electron beam to that area. The STM measurements were performed in a separate vacuum chamber with base pressure below $3\times10^{-11}$ Torr. A graphene-covered Ir(111) sample was transferred in air from the LEEM apparatus to the STM chamber and then annealed in vacuum at 900 K for 8 hours before imaging at room temperature.

## 3. Results and discussion

### *3.1 Four different rotational variants of graphene on Ir(111)*

Before presenting the three other variants of graphene on Ir(111), we report some LEEM-based characterization of the graphene variant studied by the Cologne group.[18, 19, 23, 32] Fig. 1a depicts two graphene islands as imaged by LEEM. The selected-area diffraction pattern obtained from this graphene variant is shown in Fig. 2b. The diffraction pattern has superstructure spots centered around the specular beam and the 1$^{st}$-order diffraction spots of the Ir, as evident through comparison with the LEED pattern of clean Ir(111) in Fig. 2a. This pattern from the graphene island is identical to the LEED patterns reported in Ref. [32]. We will refer to this variant of graphene, which is rotated by 0° with respect to a chosen in-plane direction of the Ir(111) surface, as "R0". As we document below, the graphene sheets in the other variants are rotated by large angles with respect to the R0 variant.

The first clue that other types of graphene occur comes from LEEM images, such as Fig. 1b. Two distinct contrasts appear in the islands of condensed-phase carbon. LEED reveals that the phase of condensed C that images "dark" corresponds to the R0 phase. The phase of condensed C that images "bright" has a different LEED pattern, as shown in Fig. 2c. Besides the diffraction spots of Ir(111) (marked by the solid green rhombus), this pattern has two 6-fold sets of spots that are both rotated by 29° ± 1° with respect to the 1$^{st}$–order Ir spots.[33] The bright spots of the "outer set", marked by the yellow-red rhombus, are ~10 % further away from the specular beam than the Ir spots. Presumably, these are the 1$^{st}$-order beams originating from the graphene lattice. The "inner set" (one group of spots is circled in Fig. 2c) is less intense and appears at positions about (0, ½) relative to the reciprocal graphene lattice. These "inner-set" beams roughly correspond to a √3×√3 superstructure with respect to Ir(111). In section 3.3 we argue that these spots result from a local coincidence of the √3×√3 superstructure of the Ir and a 2×2 superstructure of the graphene honeycomb lattice. As explained in sections 3.2 and 3.3, this type of condensed C is indeed graphene, which we label as "R30."

We find two other types of diffraction patterns of condensed-phase C on Ir(111), as shown in Fig. 2d and e. The two additional patterns are closely related to the pattern from R30 graphene. Each of the two patterns consists of the Ir spots and two 6-fold sets of spots that are both rotated by the same angle with respect to the 1$^{st}$–order Ir spots. Again, the "outer set" (marked by the dashed yellow-red rhombus) is much brighter than the "inner set". In the pattern of Fig. 2d, the two sets of beams are rotated by ~19° ± 2° from the Ir spots. As explained in section 3.4, we label this structure "R18.5". The weaker "inner set" is at positions about (0, 1/3) relative to the intense "outer set".

In the diffraction pattern shown in Fig. 2e, the two 6-fold sets are rotated by ~13.8° ± 0.7° and the "inner set" appears at about (0, 1/4)-positions. We will refer to this structure as "R14" graphene. We emphasize that the R30, R18.5, and R14 rotational variants are distinct from the small misorientations from the perfect R0 structure previously characterized by Coraux et al. (see Fig. 1 in Ref.[18]). The R14 and R18.5 have almost the same growth rates as the growth

- 4 -

rate of the R30 rotational variant described in Ref. 22. Therefore, the sizes of the all rotational domains are similar.

R30, R18.5, and R14 graphene have image contrasts in LEEM that are distinct from that of the R0 structure, as shown in Fig. 1. The origin of this contrast difference is revealed in Fig. 3, which shows the intensity of the specularly reflected electron beam as a function of electron energy for the different graphene variants. (Bright-field LEEM images are formed from the specularly reflected electron beam. Thus, the plot in Fig. 3 shows how contrast in LEEM images changes with electron energy.) Compared to R0 graphene, the other variants reflect electrons more efficiently, especially at electron energies above 7 eV. Thus, the three other variants image brighter than the R0 variant for most electron energies.

We found that the R0 variant was typically more abundant on the surface than all of the other three variants combined. For example, growth from ethylene at 1200 K covered roughly 2/3 of the surface with the R0 (dark orientation) and about 1/3 with the other (bright) variants. During graphene growth from ethylene in the temperature range from 940 K to 1200 K, the R0 graphene was typically observed to nucleate first. Then, the other variants nucleated at the edges of these R0 graphene islands. All the rotated phases grew more quickly than the R0 phase, as described in Ref. [22]. Quick growth leads to larger coverage of the rotational domains. On the other hand, sufficiently slow growth with ethylene pressure below $1\times10^{-9}$ Torr and temperature above 1400 K can result in a surface entirely covered with R0. At all growth conditions, the R14 and R18.5 phases were approximately an order of magnitude rarer than the R30 phase. We note that the non-R0 variants formed from all three C sources used – C segregated from the bulk, vapor-deposited C, and C from ethylene decomposition.

In the next section, we characterize the R30 structure using STM. Based on the STM and LEED, we then (section 3.3) propose an atomic model of this structure. We contrast the R30 structure with the more abundant R0 graphene. With this foundation, atomic models of the R18.5 and R14 structures are then presented in section 3.4.

*3.2 STM characterization of the R30 structure*

To provide structural details of the graphene sheets at the atomic level, we performed STM measurements at room temperature. Figure 4 shows a medium-scale image of the R30 structure that demonstrates that the R30 phase is as ordered as the R0 phase. Corrugations occur at two different length scales, as is typical of films uniformly strained with respect to the substrate. STM images in Fig. 5a and d directly compare the R0 structure and the R30 structure with atomic resolution. The R0 structure has a fine-scale corrugation with ~2.4 Å periodicity and a modulation of about 0.04 Å. This corrugation corresponds to the honeycomb lattice of graphene. The longer-scale corrugation occurs over about ten honeycomb spacings and has a modulation of about 0.3 Å. This long-scale corrugation of the moiré forms a hexagonal lattice that has roughly the same orientation as the honeycomb (graphene) lattice. This R0 structure has been described by N'Diaye et al.[23] as a moiré where the close-packed directions of the graphene lattice, the Ir(111) substrate, and the moiré cell are all aligned (see atomic model in Fig. 5b).

The STM topography of the R30 phase in Fig. 5d is markedly different. Here, the fine scale corrugation has a periodicity of ~5 Å and a modulation of about 0.1 Å. This corrugation produces a hexagonal lattice that is rotated ~30° with respect to the honeycomb lattice of the R0 moiré (and the surface atoms of the Ir(111) substrate). The modulation of the long-period corrugation is much less pronounced than for the R0 moiré. The long-scale corrugation has wide, and extremely shallow minima of about 0.04 Å in apparent height. These apparent depressions form a hexagonal lattice that is rotated by about 10° with respect to the substrate lattice. Only inside these shallow depressions (of the long-scale corrugation) are the dark minima of the fine-scale corrugation well-defined. Between adjacent shallow depressions, the rows of fine-scale minima shift sideways by half an inter-row distance. These shifts are readily seen by tilting the STM images in Fig. 4 and Fig. 5d and sighting down the image diagonal. From this image, the configuration of C atoms is not clear a priori. For example, it is not obvious that the configuration is consistent with defect-free graphene. In the next section, we will show that this structure is a moiré of the graphene rotated about 30° relative to the R0 structure.

### *3.3 Atomic model of the R30 structure*

In this section we develop an atomic model of the two-dimensional positions of the C atoms in the R30 structure based on the observed STM images and LEED pattern. We propose that the R30 structure of graphene forms an incommensurate overlayer, which we approximate by the periodic structure depicted in Fig. 5e: The unit cell of the graphene lattice (marked yellow-red) is rotated by 29.55° with respect to the substrate unit cell (green rhombus). (The angle of 29.55 ± 0.05° is based on the model fit. As previously reported for the R0 structure,[18] it is likely that small variations from this rotation angle occur across the substrate. These variations are below the accuracy of the LEED measurements.) The length of the graphene unit cell $h$ is equal to 0.906 × the length of the Ir cell ($a = b =$ Ir nearest-neighbor distance). This results in a ($\sqrt{124} \times \sqrt{124}$)-R9° moiré cell (marked white) whose lattice vectors are rotated 9° from the close-packed Ir directions (lattice vectors **a** and **b**). Compared to the reported dimensions [23] of the R0 structure, the graphene lattice is expanded by 0.3% (, which is probably within the margin of error).

The graphene atoms are colored according to their location relative to the underlying Ir atoms. For example, C atoms atop Ir atoms are colored red and C atoms in 3-fold hollow site are colored blue. This coloring reveals a prominent feature of the R30 structure – close to the corners of the moiré unit cell, a (2×2) superstructure of the graphene honeycomb lattice (marked by yellow hexagons) roughly coincides with the $\sqrt{3}\times\sqrt{3}$ lattice (the blue rhombus) of the substrate. That is, the centers of the graphene honeycombs are nearly centered over the Ir atoms near the corners of the marked moiré cell. For conciseness, we will refer to this near-alignment of the C atoms and the Ir atoms as an "approximate coincidence lattice". (See also section 3.4.) This local coincidence lattice presumably gives rise to the array of fine-scale minima observed with STM (black spots with ~5 Å separation in Fig. 5d) and to the weak $\sqrt{3}\times\sqrt{3}$-R~30º diffraction spots in the LEED pattern in Fig. 2c. This assignment of local topography is consistent with the model of N'Diaye et al. [32] for the R0 structure, which was developed from atomic-resolution STM images. In their model, depressed regions of the moiré occur where the centers of the graphene honeycombs are nearly centered over the Ir atoms.

To give credence to this proposed R30 structure, we generated an image based on it that mimics the STM images. We first describe the simulation procedure and show that it can produce the essential STM features of the known R0 structure. (While the image simulation is based on geometric rules, its corrugation cannot not be taken as the actual geometric corrugation.) We then compare the simulated and actual STM images of the R30 structure. In the simulations, the substrate is represented by a two-dimensional sinusoidal corrugation. A graphene lattice is placed onto the substrate, in which the C atoms are represented by filled gray circles. Their brightness is determined by the local height of the (sinusoidal) substrate. To account for the intra-layer bonding, the brightness of a given atom is computed as the average of its brightness and the mean brightness of its three nearest-neighbor (NN) atoms. NN atoms are connected by lines shaded according to the average brightness of the two neighboring atoms. The inside of each 6-membered ring in the honeycomb lattice is shaded according to the average brightness of the six C atoms of the ring and a constant offset that lowers the honeycomb centers relative to the C atoms in the graphene lattice. Finally, the computed image is smoothened by a low-pass filter.

We tested this procedure on the moiré of the known R0 graphene phase. The simulated image, Fig. 5c, reproduces well the actual STM image, Fig. 5a. Specifically, broad apparent depressions (dark) occur at the corners of the moiré unit cell. Comparison to the atomic model shows that the graphene lies close to the substrate at the cell corners because the C atoms lie near hollow sites of the Ir lattice. Away from the cell corners, C atoms lie near top sites, and appear brighter.

The image simulated for the rotated moiré of the R30 structure is shown in Fig. 5f. Two length scales dominate the simulated image. Near the corners of the ($\sqrt{124} \times \sqrt{124}$)-R9° moiré cell, the finer length-scale modulation is most pronounced. This fine-scale modulation forms an array of dark minima ("holes") with ~5 Å spacing. This array is aligned with the $\sqrt{3}$ directions of the Ir lattice. Comparison to the atomic structure (Fig. 5e) shows that these dark "holes" arise where the C atoms sit close to hollow sites in the Ir lattice, exposing the underlying Ir atoms near the center of the C rings. The long-scale modulation repeats over the moiré cell and corresponds to the very shallow modulations observed in the STM images of Fig. 4 and Fig. 5d. Away from the unit-cell corners, bright regions arranged in a defective pattern along $\sqrt{3}$ Ir directions dominate the simulated and actual images. Here C atoms are close to top sites, just as in the R0 moiré (Fig. 5b and c). The simulated image Fig. 5f also produces the observed laterals shifts in the minima of the fine-scale modulation (the "holes") by half the row separation when going along a $\sqrt{3}$ Ir direction. Given this level of agreement, we believe that the atomic model in Fig. 5e provides a reasonably accurate description of the R30 structure. Obviously, the model can be improved using more rigorous image simulations based on electronic-structure calculations.

### *3.4 Atomic models of the R18.5 and R14 structures*

We have not yet imaged the R18.5 and R14 structures with high resolution by STM. Given the rarity of these phases, it is not unexpected that they are difficult to locate. Still, based on the rotations of the graphene sheets relative to the Ir lattice found in the LEED patterns and the knowledge gained from the R0 and R30 structures, we are able to propose the atomic models of the R18.5 and R14 structures shown in Fig. 6a and b. As for R0 and R30, we assume that R18.5 and R14 graphene form incommensurate overlayers, which we approximate for simplicity by the periodic structures depicted in Fig. 6 and b. First we discuss the R18.5 structure. From the rotation of the two 6-fold sets of LEED beams in Fig. 2d (as discussed in section 3.1), we infer that the unit cell of the graphene layer (marked by yellow-red rhombus in Fig. 6a) is rotated by ~19° ± 2° with respect to the unit cell of the Ir(111) substrate (marked green). There are two equivalent rotational domains, which are both observed in LEED. Fig. 6a shows the counterclockwise rotated variant. As in the R30 structure, we constrain the length of the graphene unit cell to be 0.906 times the nearest-neighbor distance of Ir, i.e., $h = 0.906 \times a$. The superposition of a graphene layer rotated 18.45° relative to the Ir substrate results in a moiré pattern that is nearly commensurate. The exact size and rotation of the moiré unit cell (white rhombus) can only be estimated given the lack of STM data. As already seen in the R30 structure (section 3.3), the R18.5 structure also features an "approximate coincidence lattice" close to the corners of the moiré cell. Here, a 3×3 superstructure of the graphene honeycomb lattice (yellow hexagons) roughly coincides with a $\sqrt{7} \times \sqrt{7}$ superstructure (blue rhombus) of the substrate. (Starting from an Ir atom centered below a honeycomb ring, moving one substrate unit-cell spacing in the **a** direction and two spacings in the **b** direction, one arrives at an Ir atom that is again roughly centered below a honeycomb ring.) This local coincidence lattice explains the occurrence of the 6-fold set of weak LEED spots at (0, 1/3) positions (see circle in Fig. 2d).

The same ideas straightforwardly apply to the R14 structure. The atomic model in Fig. 6b shows the clockwise-rotated variant of R14 graphene. (Counterclockwise-rotated domains occur with equal probability.) The rotation angle of 13.9° between the graphene unit cell (yellow-red rhombus) and the Ir(111) unit cell (green rhombus) corresponds to the angle between the Ir spots and the graphene spots of Fig. 2e. The weak graphene spots at graphene-(0,1/4) positions (see circle in Fig. 2e) correspond to a coincidence lattice between a 4×4 superstructure of the graphene honeycomb (yellow hexagons) and a $\sqrt{13} \times \sqrt{13}$ superstructure (blue rhombus) of the substrate. The base vector of the unit cell of the coincidence lattice (blue in Fig. 6b) is equal to the sum **b** + 3**a**. For a ratio of $h/a = \sqrt{13}/4 = 0.9014$, this coincidence lattice would extend across the entire surface and its unit cell would be identical to the moiré cell (marked by the white rhombus). N'Diaye et al.[23] measured an almost identical value $h/a = 0.903$ for the R0 structure. This agreement suggests that the simple commensurate structural model of Fig. 6b is close to the actual structure.

We end this section by observing that the four rotational variants of graphene on Ir(111) can be described and classified by simple geometric rules. We again note that all four rotational variants, R0, R30, R18.5 and R14, have approximate coincidence lattices close to the corners of their moiré cells. The substrate and graphene form an approximate coincidence lattice if multiples of their unit-cell vectors roughly match. In the R0 structure, the periodicity of the coincidence lattice, the Ir lattice and the graphene lattice are about the same (Fig. 5b). In the R30 (Fig. 5e), R18.5 (Fig. 6a), and R14 (Fig. 6b) structures, the periodicity of the coincidence lattices are about two, three, and four times the graphene-cell lengths. More precisely, all four observed graphene structures have in common that the base vectors $\mathbf{v_a}$ and $\mathbf{v_b}$ of the coincidence lattices can be approximated by a sum of multiples of the Ir lattice vectors (**a** and **b**), as well as by multiples of the graphene lattice vector **h**. For the R0 structure we find $\mathbf{v_a} \approx \mathbf{h} \approx \mathbf{a}$, for R30 $\mathbf{v_a} \approx 2\mathbf{h} \approx \mathbf{a}+\mathbf{b}$, for R18.5 $\mathbf{v_a} \approx 3\mathbf{h} \approx 2\mathbf{a}+\mathbf{b}$, and for R14 $\mathbf{v_a} \approx 4\mathbf{h} \approx 3\mathbf{a}+\mathbf{b}$. Apparently it is energetically or kinetically favorable to form these approximate coincidence lattices that have short periods. Such simple geometric rules of constructing these coincidence lattices might determine the rotation angles that prevail in the grown graphene sheets.

### *3.5 Orientation selection of graphene*

Why are four different orientations observed? Only one orientation can have the lowest energy, so clearly some kinetic mechanism is responsible for their existence. One hypothesis would be that the four orientations are local minima in the energy per unit area of the graphene and that the sheets relax into, and are trapped in these minima after nucleation in a continuous range of orientations. Another possibility is that the nucleation event itself leads to the orientation selection and that the orientational dependence of the graphene energy is negligible or too small to allow subsequent equilibration. The second possibility seems slightly more plausible. Pure moiré structures, in which neither substrate nor film relax and in which the energy of each film atom depends sinusoidally on its position with respect to the substrate, have energies that do not depend on orientation.[34] Since the graphene/Ir interaction has been calculated to be very small,[23] consistent with the small elastic relaxations (corrugations) observed, this could be a good model for graphene/Ir and would imply that nucleation selects the orientation.

The very rare nucleation of the rotated phases suggests that nucleation is heterogeneous – caused by rare defects in the R0 phases or perhaps by a small density of impurities. These nucleated finite-sized moiré domains can have orientational selection depending on their shape.[34] Thus, the critical nuclei might naturally have an orientational anisotropy even if the infinite moirés did not. The lack of orientation domains on Ru(0001) could then be explained by the fact that graphene interacts more strongly with Ru, which causes larger corrugations in the graphene.[3] This strong interaction could give rise to a much stronger orientational dependence of the energy of the infinite moiré, which could override a possible orientation selection at the nucleation stage. On the other hand, it is also known that elastic relaxations can select high-order commensurate structures,[34, 35] such as those we observe on Ir(111). So perhaps the relaxations, although small, are still important. Detailed energetic calculations are necessary to resolve these issues.

### *3.6 Formation of ridges during cooling*

We end this report by showing another type of defect that can occur in graphene films on Ir(111) – graphene can locally delaminate upon cooling from the synthesis temperature. Figure 7a shows a LEEM image at the synthesis temperature (1200 K) of a nearly complete layer of graphene. The only noticeable boundaries in the film occur between the majority orientation of graphene (dark) and the minority orientations (bright). After cooling to room temperature, Fig. 7b shows that the film contains numerous new linear features. These linear features cross the boundaries between majority and minority orientations of graphene. Although less obvious, Fig. 1c also shows these features in the partially complete graphene film at room temperature.

STM provides insight into the nature of these linear features observed with LEEM. The image in Fig. 7c,d shows a straight ridge that extends about 0.55 nm above the adjacent surface covered with a R0 graphene domain. Similar ridges have been observed in graphene films on SiC after cooling from the decomposition temperature.[36,37] The large thermal contraction of the SiC substrate relative to the graphene film during cooling was believed to be responsible for the ridges. Ir and graphene also have a large difference in thermal expansion. For example, Ir contracts by about 0.8% from 1000°C to room temperature[38] while the in-plane dimensions of graphite hardly change over this same temperature range.[39] Our observation that the ridges only formed upon cooling provide further support for the interpretation that the ridges result from the well-studied phenomenon of film buckling under compression.[18, 19, 23, 24, 32] The three parallel nearly horizontal linear defects in the STM image of Fig. 4 might also be a manifestation of such buckling.

## 4. Summary and conclusions

We find that graphene occurs as at least four structures on Ir(111). The four structures differ in the azimuthal orientation of the graphene sheet relative to the Ir(111) surface. Using STM we propose an atomic model of the variant that is rotated ~30° relative to the previously characterized, most-abundant variant. We also propose atomic models of the other two variants, which are rotated ~14° and ~18.5°. All four variants are moiré structures. They can be classified according to a quasi-periodic motif, whose size equals multiples of 1, 2, 3, and 4 of the unit-cell lengths of the graphene lattice. We also observe that cooling graphene from the synthesis temperature can introduce networks of ridges where the film has locally delaminated.

**Note added in proof:** Similar formation of ridges in graphene on Ir(111) has been recently reported by Alpha T. N'Diaye, Raoul van Gastel, Antonio J. Martinez-Galera, Johann Coraux, Hichem Hattab, Dirk Wall, Frank-J. Meyer zu Heringdorf, Michael Horn-von Hoegen, Jose M. Gomez-Rodriguez, Bene Poelsema, Carsten Busse, and Thomas Michely (Relaxation of compressive strain in epitaxial graphene through wrinkle formation, arXiv:0906.0896 <http://arxiv.org/abs/0906.0896>.

**Acknowledgments**

The authors thank P. J. Feibelman for helpful discussion. This work was supported by the Office of Basic Energy Sciences, Division of Materials Sciences and Engineering of the US DOE under Contract No. DE-AC04-94AL85000.

**References**

[1] A. K. Geim, and K. S. Novoselov, Nature Materials **6**, 183 (2007).

[2] K. S. Novoselov, A. K. Geim, S. V. Morozov, D. Jiang, Y. Zhang, S. V. Dubonos, I. V. Grigorieva, and A. A. Firsov, Science **306**, 666 (2004).

[3] J. Wintterlin, and M. L. Bocquet, Surface Science **603**, 1841 (2009).

[4] Y. S. Dedkov, M. Fonin, U. Rudiger, and C. Laubschat, Physical Review Letters **100**, 107602 (2008).

[5] M. Eizenberg, and J. M. Blakely, Surface Science **82**, 228 (1979).

[6] Y. Gamo, A. Nagashima, M. Wakabayashi, M. Terai, and C. Oshima, Surface Science **374**, 61 (1997).

[7] J. C. Shelton, H. R. Patil, and J. M. Blakely, Surface Science **43**, 493 (1974).

[8] A. L. V. de Parga, F. Calleja, B. Borca, M. C. G. Passeggi, J. J. Hinarejos, F. Guinea, and R. Miranda, Physical Review Letters **100**, 056807 (2008).

[9] P. J. Feibelman, Surface Science **103**, L149 (1981).

[10] J. T. Grant, and T. W. Haas, Surface Science **21**, 76 (1970).

[11] F. J. Himpsel, K. Christmann, P. Heimann, D. E. Eastman, and P. J. Feibelman, Surface Science **115**, L159 (1982).

[12] J. Hrbek, J. Vac. Sci. Tech. A **4**, 86 (1986).

[13] E. Loginova, N. C. Bartelt, P. J. Feibelman, and K. F. McCarty, New J. Phys. **10**, 093026 (2008).

[14] S. Marchini, S. Gunther, and J. Wintterlin, Physical Review B **76**, 075429 (2007).

[15] D. Martoccia, P. R. Willmott, T. Brugger, M. Bjorck, S. Gunther, C. M. Schleputz, A. Cervellino, S. A. Pauli, B. D. Patterson, S. Marchini, J. Wintterlin, W. Moritz, and T. Greber, Phys. Rev. Lett. **101**, 126102 (2008).

[16] P. W. Sutter, J.-I. Flege, and E. A. Sutter, Nature Materials **7**, 406 (2008).

[17] M. C. Wu, Q. Xu, and D. W. Goodman, Journal of Physical Chemistry **98**, 5104 (1994).

[18] J. Coraux, A. T. N'Diaye, C. Busse, and T. Michely, Nano Letters **8**, 565 (2008).

[19] J. Coraux, A. T. N'Diaye, M. Engler, C. Busse, D. Wall, N. Buckanie, F. J. M. Z. Heringdorf, R. van Gastei, B. Poelsema, and T. Michely, New J. Phys. **11**, 023006 (2009).

[20] N. R. Gall, E. V. Rut'kov, and A. Y. Tontegode, Physics of the Solid State **46**, 371 (2004).

[21] N. R. Gall', E. V. Rut'kov, and A. Y. Tontegode, Carbon **38**, 663 (2000).

[22] E. Loginova, N. C. Bartelt, P. J. Feibelman, and K. F. McCarty, New Journal of Physics **11**, 063046 (2009).

[23] A. T. NDiaye, S. Bleikamp, P. J. Feibelman, and T. Michely, Physical Review Letters **97**, 215501 (2006).

[24] A. T. NDiaye, S. Bleikamp, P. J. Feibelman, and T. Michely, Physical Review Letters **101**, 219904 (2008).

[25] M. Enachescu, D. Schleef, D. F. Ogletree, and M. Salmeron, Physical Review B **60**, 16913 (1999).

[26] J. C. Hamilton, and J. M. Blakely, J. Vac. Sci. Tech. **15**, 559 (1978).

[27] Z. P. Hu, D. F. Ogletree, M. A. Van Hove, and G. A. Somorjai, Surface Science **180**, 433 (1987).

[28] T. A. Land, T. Michely, R. J. Behm, J. C. Hemminger, and G. Comsa, Surface Science **264**, 261 (1992).

[29]H. B. Lyon, and G. A. Somorjai, Journal of Chemical Physics **46**, 2539 (1967).

[30]J. W. May, Surface Science **17**, 267 (1969).

[31]M. Sasaki, Y. Yamada, Y. Ogiwara, S. Yagyu, and S. Yamamoto, Phys. Rev. B **61**, 15653 (2000).

[32]A. T. N'Diaye, J. Coraux, T. N. Plasa, C. Busse, and T. Michely, New Journal of Physics **10**, 043033 (2008).

[33] Statistics for the rotations of the R30, R18.5, and R14 variants are based on averaging 14, 22, and 12 LEED patterns, respectively.

[34]F. Grey, and J. Bohr, Europhysics Letters **18**, 717 (1992).

[35]C. R. Fuselier, J. C. Raich, and N. S. Gillis, Surface Science **92**, 667 (1980).

[36]Z. G. Cambaz, G. Yushin, S. Osswald, V. Mochalin, and Y. Goyotsi, Carbon **46**, 841 (2008).

[37]J. J. Halvorson, and R. T. Wimber, Journal of Applied Physics **43**, 2519 (1972).

[38]J. B. Nelson, and D. P. Riley, Proc. Phys. Soc. London **57**, 477 (1945).

[39]L. B. Freund, and S. Suresh, *Thin Film Materials* (Cambridge University Press, Cambridge, 2003), p. Chap. 5.

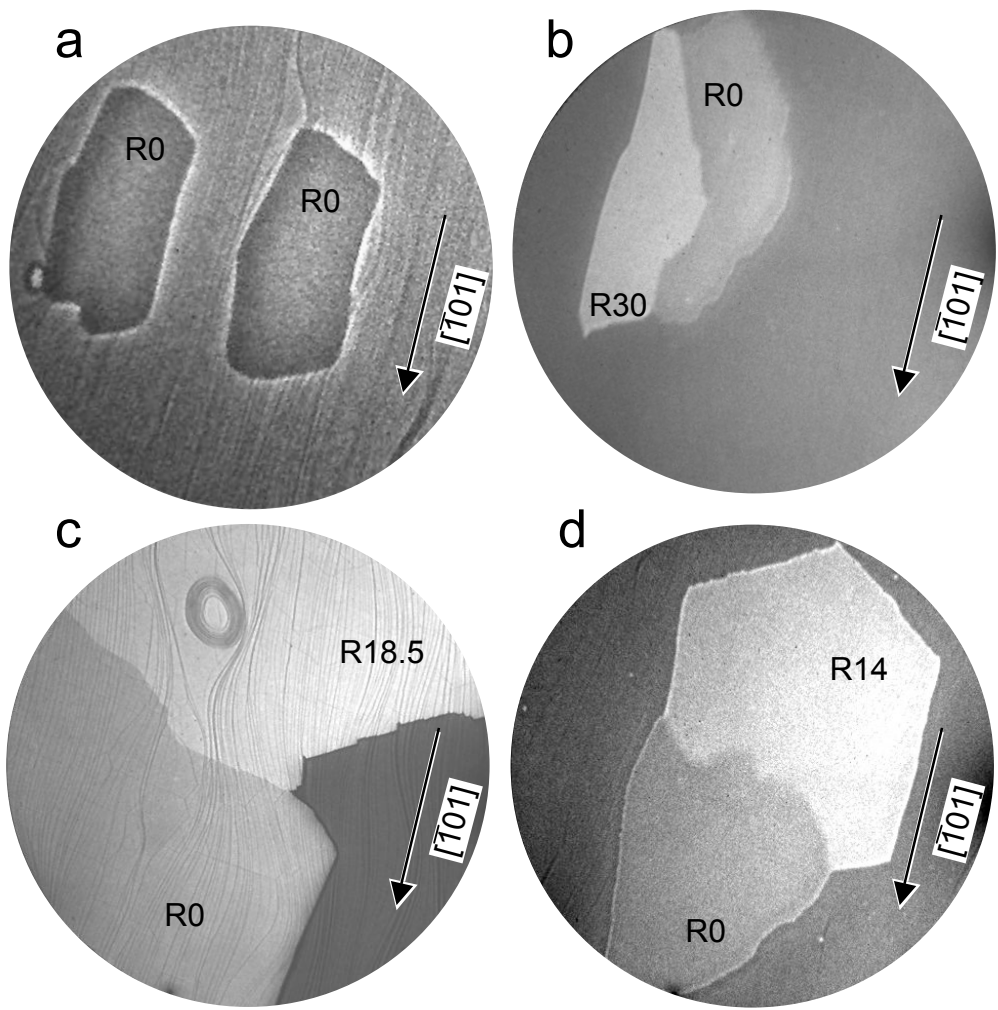

**Figure 1.** LEEM images of (a) (20 μm field-of-view) R0 graphene islands (dark) grown on Ir(111) by ethylene deposition at 1320 K and $C_2H_4$ pressure $3 \times 10^{-9}$ Torr; (b) (7 μm field-of-view) R0 (right, dark) and R30 (left, bright) graphene islands grown by ethylene at 1100 K; (c) (20 μm field-of-view) R0 (bottom, dark) and R18.5 (top, bright) types of graphene grown at 1200 K (LEEM is taken at 300 K); (d) (20 μm field-of-view) R0 (bottom, dark) and R14 (top bright) structures of graphene grown at 1100 K. Unlabeled regions are Ir(111). Arrows show Ir directions.

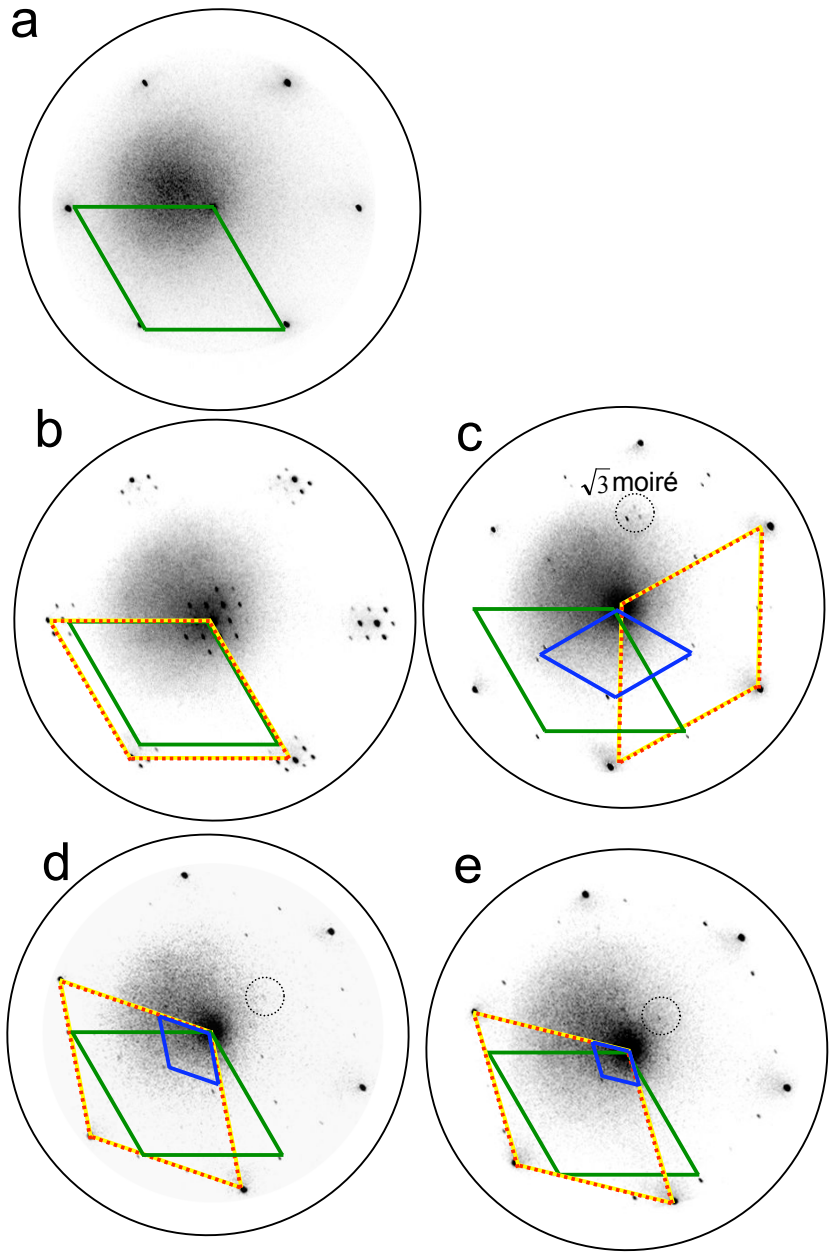

**Figure 2.** LEED images of clean Ir(111) (a), R0 (b), R30 (c), R18.5 (d), and R14 (e) graphene taken at 40 eV. Unit cells of the reciprocal Ir(111) lattice (green), honeycomb carbon (dashed yellow-red), and approximate coincidence lattice (blue) are shown.

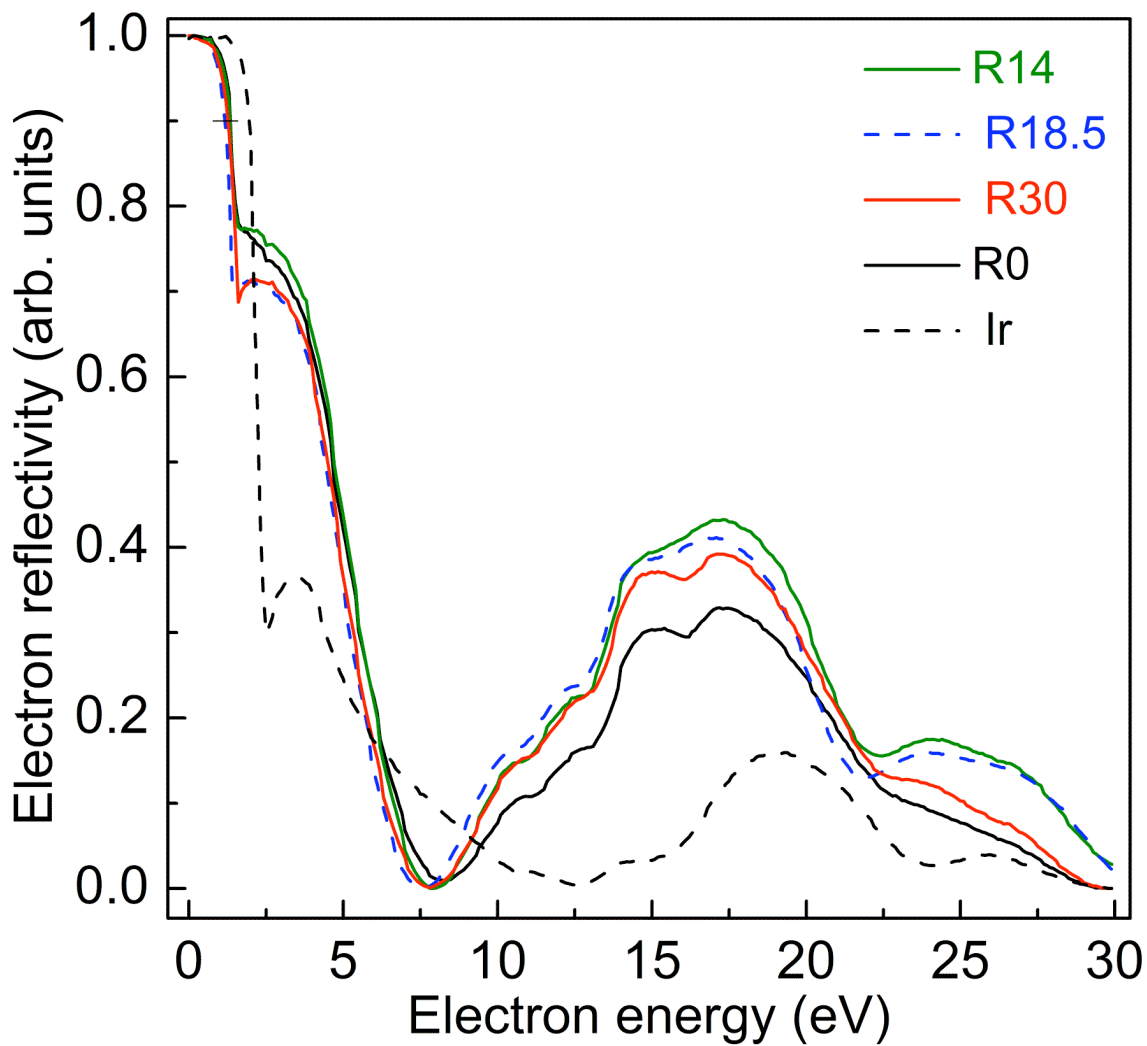

**Figure 3.** Electron reflectivity versus electron energy of clean Ir(111) (dashed black), R0 (solid black), R30 (solid red), R18.5 (dashed blue), and R14 (solid green) graphene domains taken at 300 K. Graphene was grown by ethylene deposition at 1200 K and $C_2H_4$ pressure ~5 × $10^{-8}$ Torr. The orientations of the graphene domains from which electron reflectivities were measured from LEEM images were determined by LEED.

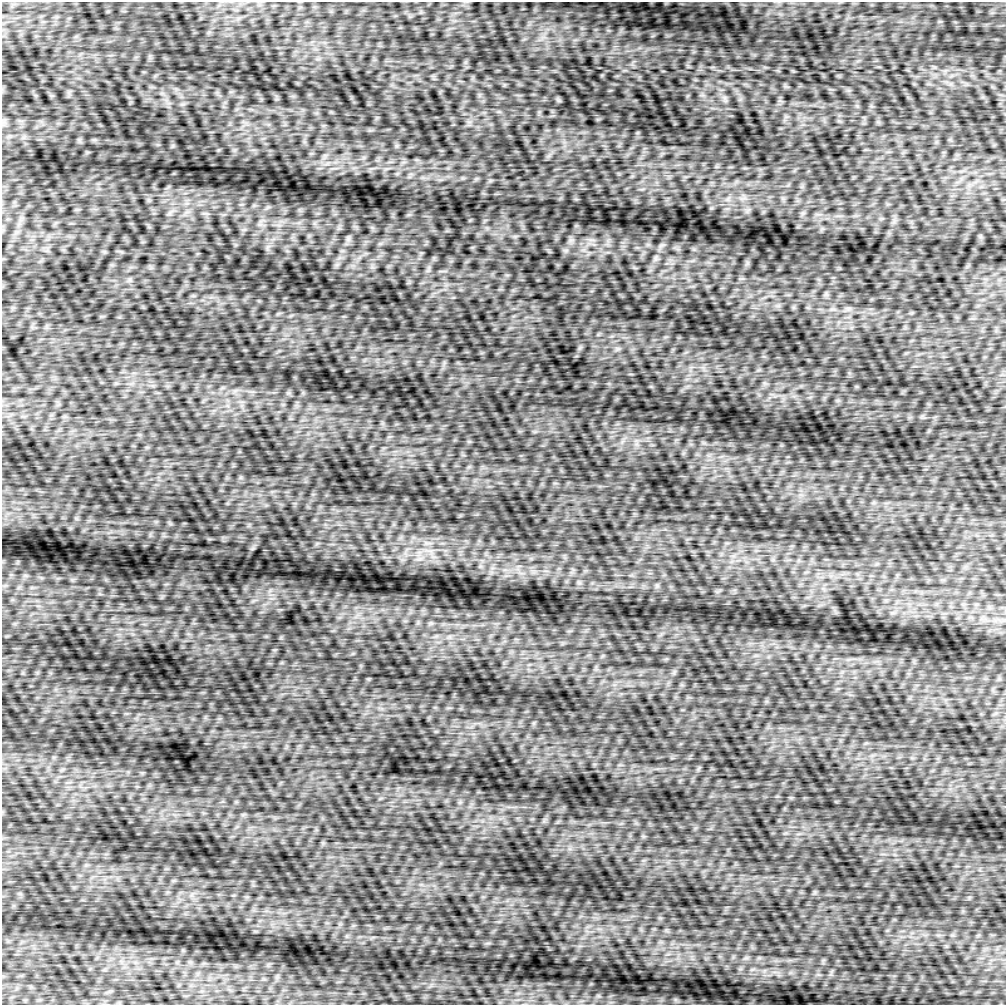

**Figure 4.** STM image of the R30 structure of graphene on Ir(111) (40 nm × 40 nm, Vs = 0.2 V, It = 60 nA).

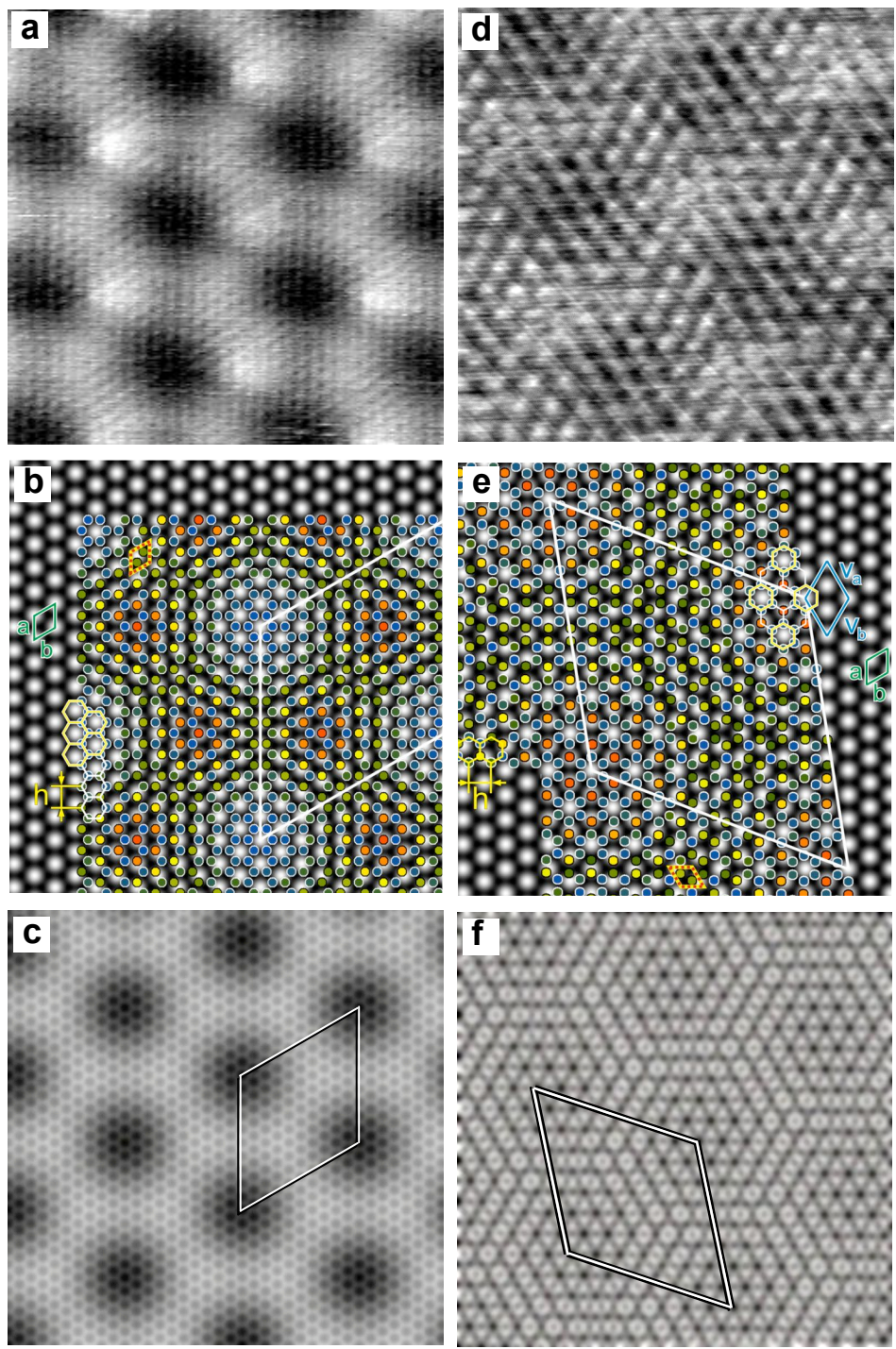

**Figure 5.** STM images (8 nm × 8 nm) of R0 (a) ($V_s$ = 0.3 V, $I_t$ = 50 nA) and R30 (d) ($V_s$ = 0.2 V, $I_t$ = 50 nA) graphene islands on Ir(111); schematic representation (48 Å × 48 Å) of Ir(111) and graphene layer rotated by 0 degrees (b) and 30 degrees (e) with respect to the substrate. Unit cells of Ir(111) substrate (green), carbon honeycomb lattice (dashed yellow-red), approximate coincidence lattice (blue), and moiré (white) are shown with colored rhombuses. Schematic illustrations (8 nm × 8 nm) of the graphene layer rotated by 0 degrees (c) and 30 degrees (f) on Ir(111), in which apparent height of the graphene atoms is indicated by brightness. Graphene is represented as a honeycomb mesh.

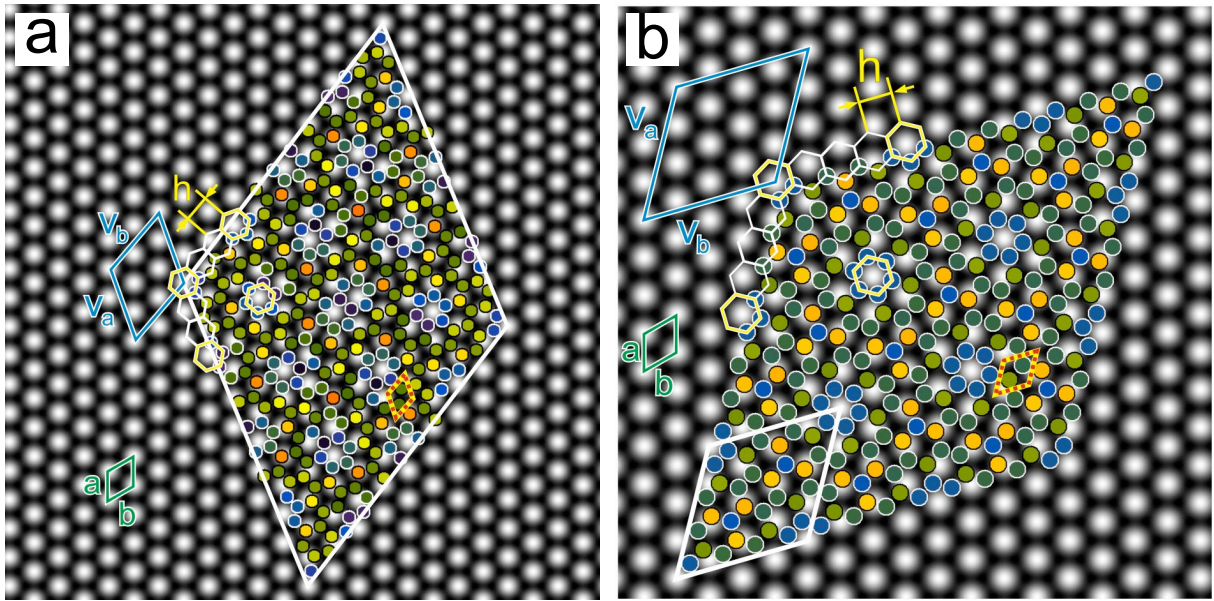

**Figure 6.** Atomic models of Ir(111) and graphene layer rotated by 18.5 degrees (56 Å × 56 Å) (a) and 14 degrees (42 Å × 42 Å) (b) with respect to the Ir substrate. Shown are unit cells of

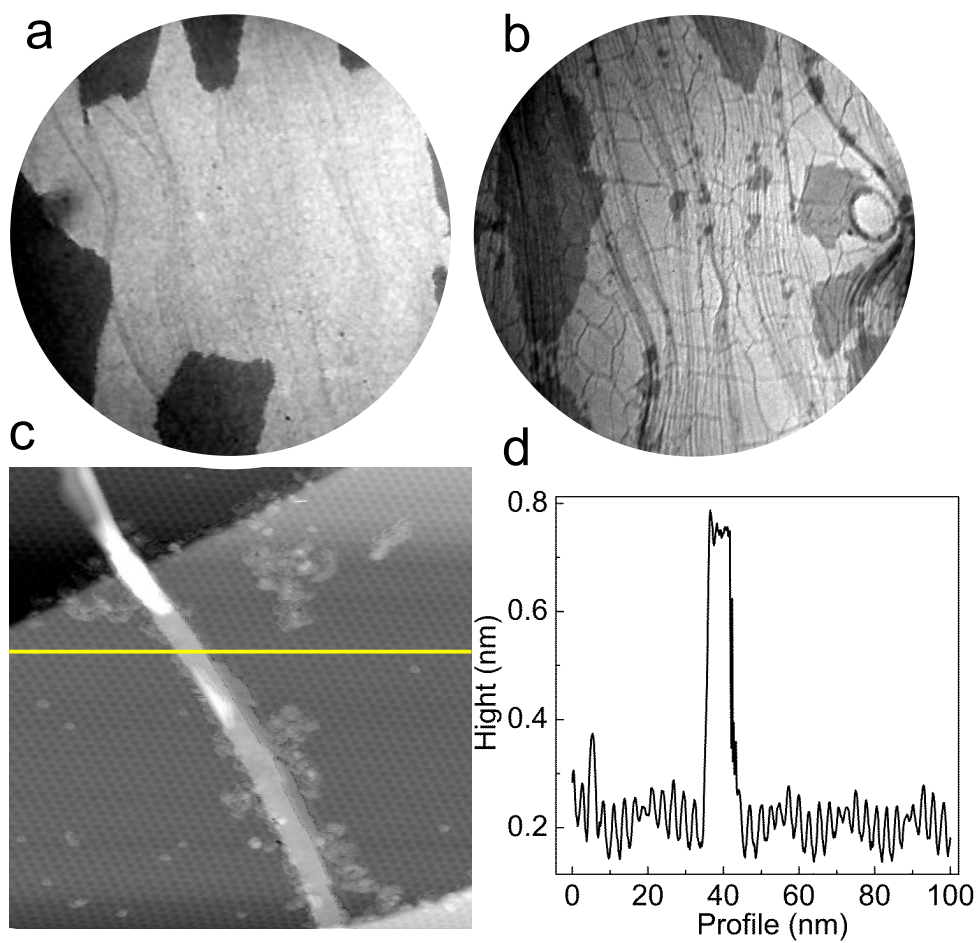

**Figure 7.** LEEM images of 20 mm field-of-view of the Ir(111) surface totally covered with R0 (dark regions) and R14 (bright regions) graphene at 1200 K (a) and 300 K (b); (c) STM image (100 nm ´ 100 nm) of the R0 graphene-covered Ir(111) at 300 K and (d) profile along the yellow line shown in (c).